\documentclass[proof]{WileyASNA-v1}

\articletype{Article Type}%

\received{xx xxxx xxxx}
\revised{xx xxxx xxxx}
\accepted{xx xxxx xxxx}

\begin{document}

\title{Tracing X-ray and H{\sc I} absorption in peaked spectrum sources}

\author[1,2]{Emily F. Kerrison*}

\author[2,1]{Vanessa A. Moss}

\author[1,2,3]{Elaine M. Sadler}

\author[3,4]{James R. Allison}

\authormark{Emily F. Kerrison \textsc{et al}}

\address[1]{\orgdiv{Sydney Institute for Astronomy, School of Physics A28}, \orgname{University of Sydney}, \orgaddress{\state{NSW}, \country{Australia}}}

\address[2]{\orgdiv{CSIRO Space \& Astronomy}, \orgname{ATNF}, \orgaddress{\state{NSW, 1710}, \country{Australia}}}

\address[3]{ARC Centre of Excellence for All-Sky Astrophysics in 3 Dimensions (ASTRO 3D)}

\address[4]{\orgdiv{Sub-Department of Astrophysics, Department of Physics}, \orgname{University of Oxford}, \orgaddress{Denys Wilkinson Building, Keble Rd., \state{Oxford OX1 3RH}, \country{UK}}}

\corres{*Emily Kerrison, \email{eker0753@uni.sydney.edu.au}}

\presentaddress{This is sample for present address text this is sample for present address text}

\abstract{Recent studies have shown that both 21cm \mbox{H\,{\sc I}} absorption and soft X-ray absorption serve as excellent tracers of the dense and dusty gas near the active nucleus of young radio galaxies, offering new insight into the physical nature of the circumnuclear medium. To date, a correlation between the column densities derived using these absorption processes has been observed within Compact Steep Spectrum (CSS) and Gigahertz-Peaked Spectrum (GPS) radio sources. While it is possible that this correlation exists within the broader radio population, many samples of radio galaxies make this difficult to test due to selection effects. This paper explores the possibility of a correlation in the broader radio population by analysing a historic sample of 168 radio sources compiled from the literature in such a way so as to minimise selection bias. From this historic sample we conclude that there is some evidence for a correlation between \mbox{H\,{\sc I}} and soft X-ray absorption outside of peaked spectrum sources, but that the selection bias towards these sources makes further analysis difficult using current samples. We discuss this in the context of the SEAFOG project and how forthcoming data will change the landscape of future absorption studies.}

\jnlcitation{\cname{%
\author{Kerrison E.F.}, 
\author{Moss V.A.},
\author{Sadler E.M.}, and
\author{Allison J.R.}} (\cyear{2021}), 
\ctitle{Tracing X-ray and \mbox{H\,{\sc I}} absorption in peaked spectrum sources}, 
\cjournal{Astronomiche Nachrichten}, \cvol{xx;xx:x--x}.}

\maketitle

\footnotetext{\textbf{Abbreviations:} GPS, Gigahertz-Peaked Spectrum; CSS, Compact Steep Spectrum; PS, Peaked Spectrum}

\section{Introduction}\label{sec1}

Multi-wavelength studies of radio galaxies have revealed that the dense and dusty gas near the active nucleus can be probed by both soft X-ray absorption and 21cm \mbox{H\,{\sc I}} absorption, providing new insight into the physical nature of this gaseous environment \citep[e.g.][]{Vink2006, Glowacki2017, Moss2017}. Each absorption measure can be used independently to derive a column density of hydrogen gas along the line of sight, and so comparing the two may inform us whether multi-wavelength emissions trace co-located gas or not. Many of the studies interested in the connection between radio and X-ray emission have selected samples comprising exclusively peaked spectrum (PS) sources because of their compact nature and young evolutionary state \citep[e.g.][]{Siemiginowska2016, Ostorero2016}, while those studies which have not tend to inherit the selection effects of previous samples. This has led to an over-representation of peaked spectrum sources in the majority of cases.

The tendency to favour PS sources within absorption studies means the relationship between \mbox{H\,{\sc I}} and soft X-ray absorption is relatively well-characterised within this subset of radio sources. However, studying these potentially skewed samples does make it difficult to relate properties of PS sources to the broader radio population, and thus we cannot say for sure whether they are unique in providing an environment conducive to both soft X-ray and \mbox{H\,{\sc I}} absorption. 

We begin to address this question of the uniqueness of PS sources by examining a sample of 168 radio galaxies compiled from the literature. This sample has no explicit constraints on source morphology or SED shape, so it provides a useful insight into the selection effects which are present when drawing from previous surveys.

\subsection{Peaked spectrum sources}\label{sec1a}

We use here the term `Peaked Spectrum' (PS) to refer collectively to `Gigahertz-Peaked Spectrum' (GPS) sources, `Compact Steep Spectrum' (CSS) sources, `Megahertz Peaked Spectrum' (MPS) sources and `High Frequency Peakers' (HFP). This choice reflects the fact that all PS sources share a number of characteristics. In particular, all have convex radio spectra which turn over anywhere between about 100\,MHz (for CSS sources) and 10\,GHz (for HFPs), likely due to either Free-Free Absorption or Synchrotron Self-Absorption \citep{Bicknell1997, Snellen2000}. Further, the ongoing discussion surrounding the mechanism of this turnover has made many PS sources targets for detailed observation and modelling \citep{Tingay2015}. Finally, all PS sources are morphologically compact at Gigahertz frequencies ($\lesssim 20$ kpc across), and to date many of those observed are radio bright \citep{ODea2021}.

\subsection{HI Absorption}\label{sec:hIabsorption}

The \mbox{H\,{\sc I}} 21-cm absorption line tends to trace cold ($\text{T}_{s} \approx$ 100K) gas well below the temperatures required for ionisation. It is possible that in PS sources, this cold gas is also responsible (at least in part) for the spectral turnover at low frequencies.

Our ability to detect \mbox{H\,{\sc I}} absorption of a given column density and spin temperature is dependent only on the flux density of the absorbed background source, and the covering fraction of the absorbing gas \citep{Morganti2018}. Thus, the radio brightness and compactness of PS sources mentioned above make them ideal objects to search for \mbox{H\,{\sc I}} absorption, explaining why they often feature heavily in targeted absorption searches \citep[e.g. the recent survey of ][]{Grasha2019}. Since the depth of an absorption line combined with knowledge of $\text{T}_s$ allows us to derive the column density of neutral hydrogen ($N_{\text{HI}}$) along the sight-line, associated lines inform us about the local gaseous environment of absorbed sources.

\subsection{X-ray Absorption}

Where radio AGN are X-ray emitters (and not all are), they often exhibit a soft excess due to accretion, which may be absorbed by dense, circumnuclear material \citep{Done2012}. By modelling this absorption it is possible to determine the column density of all hydrogen ($N_\text{H}$) within the inner regions of the AGN.

To date, the PS sources studied at X-ray wavelengths have all had spectra which are well-fit by an absorbed power law, indicating that this soft X-ray absorption is occurring within their central regions \citep{Worrall2004, Siemiginowska2016}.

\subsection{From radio to X-ray: connecting absorption across the spectrum}\label{sec:radio2xray}

Since estimates of the gas column density along a sight-line can be derived from both \mbox{H\,{\sc I}} and soft X-ray absorption independently, it is interesting to consider whether these estimates ever agree, which might imply they trace co-spatial gas. Indeed \cite{Vink2006} found a correlation between the two within GPS sources, but noted that the systematically lower $N_{\text{HI}}$ estimate could be because it traces gas at larger radii. Later, \cite{Ostorero2016} found this same correlation within an independent sample of GPS sources, and argued that 
these processes do trace co-located, circumnuclear gas. Consistent with such a correlation, \cite{Glowacki2017} and \cite{Moss2017} found that the degree of \mbox{H\,{\sc I}} and soft X-ray absorption is also correlated in more diverse samples of radio sources not limited by radio SED shape. 
However, given the preference for PS sources in targeted \mbox{H\,{\sc I}} absorption searches, it is likely that even in historical unconstrained samples a disproportionately large fraction are peaked spectrum. Accordingly, we revisit here the sample used in \cite{Moss2017} to determine whether the correlation observed is due to an underlying preference for PS sources in the literature. 

Ultimately, we want to know whether this correlation in absorption exists only in PS sources, or in the wider radio population too. If it is limited to PS sources, that tells us there is something unique about these objects, either in their compactness or circumnuclear composition. If not, it opens up a new avenue for exploring their connection with the broader radio population. Either way a better understanding of the preexisting samples within the literature is necessary.

\section{A historic sample in radio and X-ray}\label{sec2}

To explore the relationship between \mbox{H\,{\sc I}} absorption, soft X-ray absorption and PS sources, we built a sample of radio sources from the literature, selecting two studies of \mbox{H\,{\sc I}} absorbers for which SED shape was not a constraint so as to minimise the selection bias of our sample towards PS sources.

Our sample is drawn from \cite{Moss2017} and \cite{Curran2010}, which are in turn compiled from a heterogeneous collection of \mbox{H\,{\sc I}} absorption searches (see references therein). Combining these two samples gave us a control, non X-ray population alongside the X-ray emitters of \cite{Moss2017}, and produced 168 unique radio sources in total as outlined in Table \ref{sourcetable}\!\!\!. The Moss sample was X-ray selected, whereas the Curran and Whiting sample was compiled as a complete list of $z \ge 0.1$ \mbox{H\,{\sc I}} absorption searches at the time of publication, so by combining them we minimise any selection bias towards X-ray bright radio sources. 

\begin{table}[ht]%
\caption{Breakdown of our historic sample.\label{sourcetable}}
\centering
\begin{tabular*}{240pt}{@{\extracolsep\fill}lccD{.}{.}{3}c@{\extracolsep\fill}}
\toprule
\textbf{Reference} & \textbf{Constraints}  & \textbf{Sources}   \\
\midrule
\cite{Moss2017} & \begin{tabular}{@{}c@{}}Counterpart in \\ 3XMM-DR4\end{tabular}  & 96 + 5 \tnote{$^a$}   \\
\cite{Curran2010} & z $\geqslant$ 0.1  &  67  \\
\bottomrule
\end{tabular*}
\begin{tablenotes}
\item$^a$ 96 sources had been searched for \mbox{H\,{\sc I}} absorption prior to \cite{Moss2017}, 5 sources were searched for the first time.
\end{tablenotes}
\end{table}

The heterogeneous nature of this sample meant we were unable to simply adopt preexisting radio classifications to determine the fraction of PS sources, as many objects had multiple classifications across the literature. Instead, we re-identified PS sources by constructing the radio SED of each object between 50\,MHz and 20\,GHz, then using least-squares fitting and a reduced-$\chi^2$ test to determine whether they were better characterised by a simple power law of the form $S \propto \nu^\alpha$, or by an absorbed power law characteristic of a PS source, as discussed in \cite{Snellen1998}. Photometry for this procedure was obtained automatically via NED object searches for each source.

\subsection{Population statistics}\label{sec2b}

\begin{figure}[ht]
\centerline{\includegraphics[width=230pt, trim={1mm 0 0 1mm},clip]{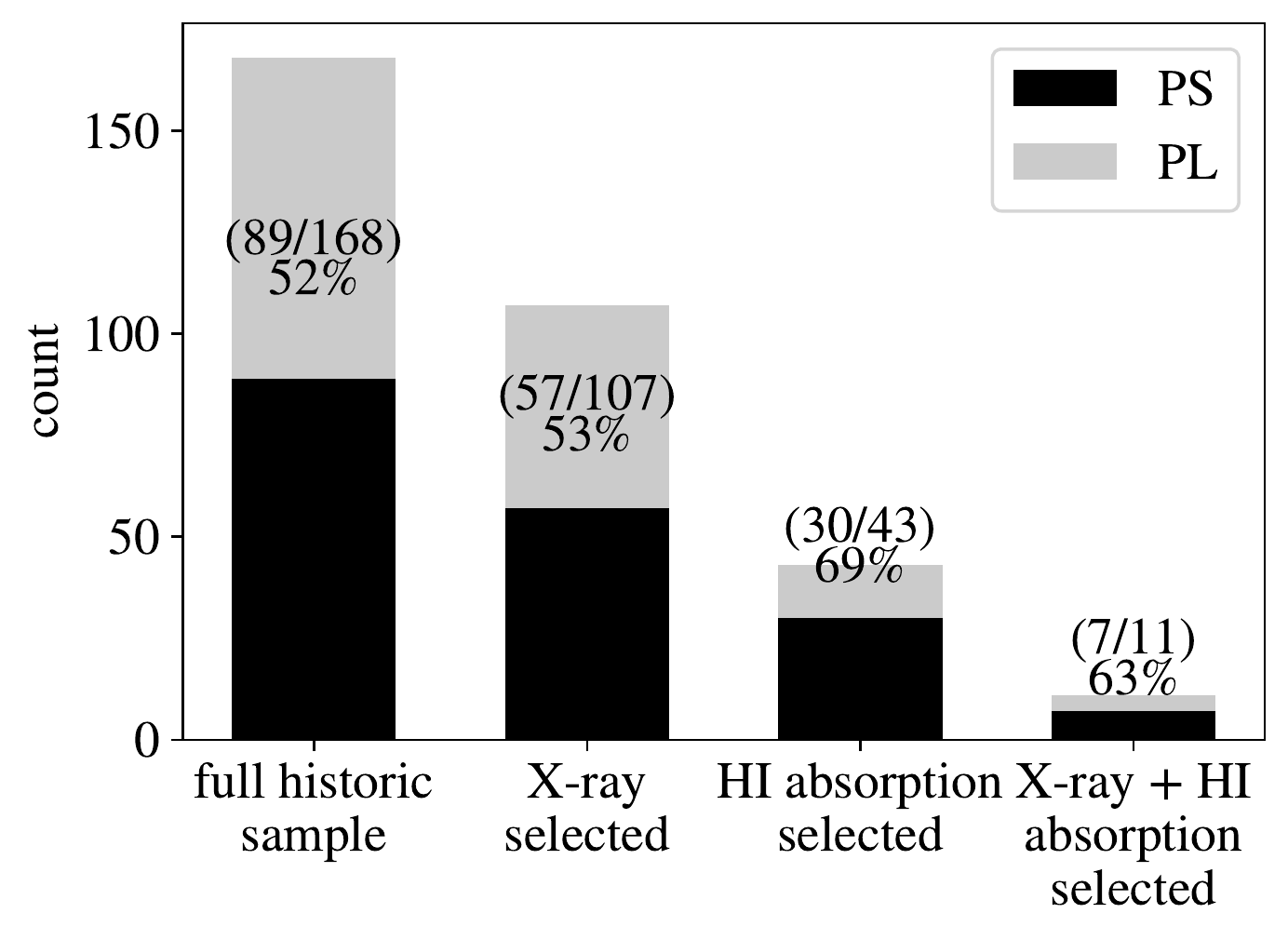}}
\caption{Proportion of PS sources within our historic sample and various subsets of the sample. The sample starts with a bias towards PS sources, and this persists in all of the subsets. \label{fig:barplot}}
\end{figure}

After classifying this historic sample according to SED shape, it was possible to consider how the correlation between 21cm \mbox{H\,{\sc I}} absorption and soft X-ray absorption changed (or not) with the degree of spectral curvature.
The proportion of our historic sample that is PS is shown in Figure \ref{fig:barplot}\!\! as a percentage of the whole, alongside the PS proportion of three subsets: the X-ray selected sources, sources with \mbox{H\,{\sc I}} absorption, and sources with both \mbox{H\,{\sc I}} and soft X-ray absorption, where soft X-ray absorption is defined as $\frac{f5}{f1} > 100$, the ratio of XMM band 5 (4.5-12keV) over band 1 (0.2-0.5keV) following \cite{Moss2017}.

This figure shows that our historic sample is still biased towards PS sources; just over half of the sample has a peaked spectrum, which is large compared to the 10-30\% noted by \cite{ODea1998} for the GHz-emitting radio population. It is even further from the 4.5\%  found by \cite{Callingham2017} for the proportion of PS sources at lower frequencies. This suggests that samples built from historic searches for \mbox{H\,{\sc I}} absorption are biased towards PS sources even when SED shape and source type are not intentionally constrained as part of sample selection. It is likely that this is a consequence of the preferential treatment of PS sources in targeted absorption surveys because of their compactness and brightness as discussed in section \ref{sec1}. 

Given this significant over-representation of PS sources in the full sample, it is interesting that they do not completely dominate the subset selected for \mbox{H\,{\sc I}} absorption, nor the X-ray emitting subset which was analysed in \cite{Moss2017}. In fact, their proportion shows only a modest increase to 69\% (30/43) within the \mbox{H\,{\sc I}} subset, and this does not differ much from the subset selected on both \mbox{H\,{\sc I}} and soft X-ray absorption where they make up 63\% (7/11). Although these last two subsets are relatively small, the fact that the fraction of PS sources does not increase between them is interesting. It may suggest the correlation between \mbox{H\,{\sc I}} and soft X-ray absorption observed in this sample by \cite{Moss2017} holds in the general radio population, and not just in PS sources. Ultimately larger samples will be required to explore this possibility further. 

\section{Summary: the future of SEAFOG}\label{conclusion}

In this work we have examined a historic sample compiled from previous \mbox{H\,{\sc I}} absorption searches as discussed in \cite{Moss2017} and \cite{Curran2010}. Even from a simple analysis of the proportions of this sample, it is clear that it is still biased towards PS sources with 52\% (89/168) classified as PS despite our choice to leave source morphology and SED shape unconstrained when constructing the sample. This bias can be explained by the preferential selection of PS sources in targeted \mbox{H\,{\sc I}} absorption searches, which leads to a disproportionate number of these sources in the literature. From here we must determine whether this selection bias matters, and if it does, what can be done about it. 

On the one hand, we see in our historic sample some evidence that the correlation between \mbox{H\,{\sc I}} and soft X-ray absorption may exist in the broader radio population, and this is despite the selection effects which preference PS sources. On the other, it is difficult to say whether the small fractional decrease in PS sources from the  \mbox{H\,{\sc I}} selected subset to the \mbox{H\,{\sc I}} and soft X-ray absorbed subset is significant owing to the small sample size once these constraints are imposed. Thus while our historic sample demonstrates that these absorption mechanisms do correlate outside PS sources, it does not tell us whether this correlation is more frequent in one or the other source type.

It is for this reason, among many others, that there is a push towards large area, untargeted surveys with the next generation of instruments, which will produce samples unhindered by the selection effects discussed here. Of particular interest to the question of \mbox{H\,{\sc I}} and soft X-ray correlation will be future studies conducted within the SEAFOG project (Studies of eROSITA and ASKAP-FLASH Obscured Galaxies) (Moss et al. \textit{in prep}). SEAFOG is a multi-wavelength project that will, for the first time, provide a sample of hundreds of galaxies searched for both radio and X-ray absorption in an entirely unbiased manner. This will be achieved through combining data from FLASH (the First Large Absorption Survey in HI) conducted with the Australian Square Kilometre Array Pathfinder (ASKAP) \citep{Yoon2020, allison_flashsurvey} with data from the eROSITA-DE all-sky survey \citep{Predehl2021}. Multi-wavelength, untargeted surveys of this kind will offer new insights into the nature of both PS sources and the relationship between \mbox{H\,{\sc I}} and soft X-ray absorption, and they will achieve this by providing extraordinarily large samples of radio galaxies with complementary multi-wavelength observations. However this future work would not be possible without historic samples like the one studied here, which provide the foundational understanding necessary for the science to come.

\section*{Acknowledgements}

 This research has made use of the NASA/IPAC Extragalactic Database (NED), which is operated by the Jet Propulsion Laboratory, California Institute of Technology, along the VizieR catalogue access tool, CDS, Strasbourg, France \citep{Vizierpaper}, and Astropy, a community-developed core Python package for astronomy \citep{AstropyCollaboration2013, AstropyCollaboration2018}.


\begin{thebibliography}{}

\bibitem [\protect \citeauthoryear {%
{Allison}%
\ \protect \BOthers {.}}{%
{Allison}%
\ \protect \BOthers {.}}{%
{\protect \APACyear {2021}}%
}]{%
allison_flashsurvey}
\APACinsertmetastar {%
allison_flashsurvey}%
\begin{APACrefauthors}%
{Allison}, J\BPBI R.%
, {Sadler}, E\BPBI M.%
, {Amaral}, A\BPBI D.%
\ et al.\end{APACrefauthors}%
\unskip\
\newblock
\APACrefYearMonthDay{2021}{{\APACmonth{10}}}{},
\newblock
\unskip
\newblock
\APACjournalVolNumPages{arXiv e-prints}{}{}{arXiv:2110.00469}.
\PrintBackRefs{\CurrentBib}

\bibitem [\protect \citeauthoryear {%
{Astropy Collaboration}%
\ \protect \BOthers {.}}{%
{Astropy Collaboration}%
\ \protect \BOthers {.}}{%
{\protect \APACyear {2018}}%
}]{%
AstropyCollaboration2018}
\APACinsertmetastar {%
AstropyCollaboration2018}%
\begin{APACrefauthors}%
{Astropy Collaboration}%
, Price-Whelan, A\BPBI M.%
, Sipőcz, B\BPBI M.%
\ et al.\end{APACrefauthors}%
\unskip\
\newblock
\APACrefYearMonthDay{2018}{}{},
\newblock
\unskip
\newblock
\APACjournalVolNumPages{AJ}{156}{3}{123}.
\newblock
\begin{APACrefDOI} \doi{10.3847/1538-3881/AABC4F} \end{APACrefDOI}
\PrintBackRefs{\CurrentBib}

\bibitem [\protect \citeauthoryear {%
{Astropy Collaboration}%
\ \protect \BOthers {.}}{%
{Astropy Collaboration}%
\ \protect \BOthers {.}}{%
{\protect \APACyear {2013}}%
}]{%
AstropyCollaboration2013}
\APACinsertmetastar {%
AstropyCollaboration2013}%
\begin{APACrefauthors}%
{Astropy Collaboration}%
, Robitaille, T\BPBI P.%
, Tollerud, E\BPBI J.%
\ et al.\end{APACrefauthors}%
\unskip\
\newblock
\APACrefYearMonthDay{2013}{}{},
\newblock
\unskip
\newblock
\APACjournalVolNumPages{A\&A}{558}{}{A33}.
\newblock
\begin{APACrefDOI} \doi{10.1051/0004-6361/201322068} \end{APACrefDOI}
\PrintBackRefs{\CurrentBib}

\bibitem [\protect \citeauthoryear {%
Bicknell%
, Dopita%
\BCBL {}\ \BBA {} O'Dea%
}{%
Bicknell%
\ \protect \BOthers {.}}{%
{\protect \APACyear {1997}}%
}]{%
Bicknell1997}
\APACinsertmetastar {%
Bicknell1997}%
\begin{APACrefauthors}%
Bicknell, G\BPBI V.%
, Dopita, M\BPBI A.%
\BCBL {}\ \BBA {} O'Dea, C\BPBI P\BPBI O.%
\end{APACrefauthors}%
\unskip\
\newblock
\APACrefYearMonthDay{1997}{}{},
\newblock
\unskip
\newblock
\APACjournalVolNumPages{ApJ}{485}{1}{112--124}.
\newblock
\begin{APACrefDOI} \doi{10.1086/304400} \end{APACrefDOI}
\PrintBackRefs{\CurrentBib}

\bibitem [\protect \citeauthoryear {%
Callingham%
\ \protect \BOthers {.}}{%
Callingham%
\ \protect \BOthers {.}}{%
{\protect \APACyear {2017}}%
}]{%
Callingham2017}
\APACinsertmetastar {%
Callingham2017}%
\begin{APACrefauthors}%
Callingham, J\BPBI R.%
, Ekers, R\BPBI D.%
, Gaensler, B\BPBI M.%
\ et al.\end{APACrefauthors}%
\unskip\
\newblock
\APACrefYearMonthDay{2017}{}{},
\newblock
\unskip
\newblock
\APACjournalVolNumPages{ApJ}{836}{2}{174}.
\newblock
\begin{APACrefDOI} \doi{10.3847/1538-4357/836/2/174} \end{APACrefDOI}
\PrintBackRefs{\CurrentBib}

\bibitem [\protect \citeauthoryear {%
Curran%
\ \BBA {} Whiting%
}{%
Curran%
\ \BBA {} Whiting%
}{%
{\protect \APACyear {2010}}%
}]{%
Curran2010}
\APACinsertmetastar {%
Curran2010}%
\begin{APACrefauthors}%
Curran, S\BPBI J.%
\BCBT {}\ \BBA {} Whiting, M\BPBI T.%
\end{APACrefauthors}%
\unskip\
\newblock
\APACrefYearMonthDay{2010}{}{},
\newblock
\unskip
\newblock
\APACjournalVolNumPages{ApJ}{712}{1}{303--317}.
\newblock
\begin{APACrefDOI} \doi{10.1088/0004-637X/712/1/303} \end{APACrefDOI}
\PrintBackRefs{\CurrentBib}

\bibitem [\protect \citeauthoryear {%
Done%
, Davis%
, Jin%
, Blaes%
\BCBL {}\ \BBA {} Ward%
}{%
Done%
\ \protect \BOthers {.}}{%
{\protect \APACyear {2012}}%
}]{%
Done2012}
\APACinsertmetastar {%
Done2012}%
\begin{APACrefauthors}%
Done, C.%
, Davis, S\BPBI W.%
, Jin, C.%
, Blaes, O.%
\BCBL {}\ \BBA {} Ward, M.%
\end{APACrefauthors}%
\unskip\
\newblock
\APACrefYearMonthDay{2012}{mar}{},
\newblock
\unskip
\newblock
\APACjournalVolNumPages{MNRAS}{420}{3}{1848--1860}.
\newblock
\begin{APACrefDOI} \doi{10.1111/J.1365-2966.2011.19779.X} \end{APACrefDOI}
\PrintBackRefs{\CurrentBib}

\bibitem [\protect \citeauthoryear {%
Glowacki%
\ \protect \BOthers {.}}{%
Glowacki%
\ \protect \BOthers {.}}{%
{\protect \APACyear {2017}}%
}]{%
Glowacki2017}
\APACinsertmetastar {%
Glowacki2017}%
\begin{APACrefauthors}%
Glowacki, M.%
, Allison, J\BPBI R.%
, Sadler, E\BPBI M.%
\ et al.\end{APACrefauthors}%
\unskip\
\newblock
\APACrefYearMonthDay{2017}{}{},
\newblock
\unskip
\newblock
\APACjournalVolNumPages{MNRAS}{467}{3}{2766--2786}.
\newblock
\begin{APACrefDOI} \doi{10.1093/MNRAS/STX214} \end{APACrefDOI}
\PrintBackRefs{\CurrentBib}

\bibitem [\protect \citeauthoryear {%
Grasha%
, Darling%
, Bolatto%
, Leroy%
\BCBL {}\ \BBA {} Stocke%
}{%
Grasha%
\ \protect \BOthers {.}}{%
{\protect \APACyear {2019}}%
}]{%
Grasha2019}
\APACinsertmetastar {%
Grasha2019}%
\begin{APACrefauthors}%
Grasha, K.%
, Darling, J.%
, Bolatto, A.%
, Leroy, A\BPBI K.%
\BCBL {}\ \BBA {} Stocke, J\BPBI T.%
\end{APACrefauthors}%
\unskip\
\newblock
\APACrefYearMonthDay{2019}{}{},
\newblock
\unskip
\newblock
\APACjournalVolNumPages{ApJS}{245}{1}{28--28}.
\newblock
\begin{APACrefDOI} \doi{10.3847/1538-4365/ab4906} \end{APACrefDOI}
\PrintBackRefs{\CurrentBib}

\bibitem [\protect \citeauthoryear {%
Morganti%
\ \BBA {} Oosterloo%
}{%
Morganti%
\ \BBA {} Oosterloo%
}{%
{\protect \APACyear {2018}}%
}]{%
Morganti2018}
\APACinsertmetastar {%
Morganti2018}%
\begin{APACrefauthors}%
Morganti, R.%
\BCBT {}\ \BBA {} Oosterloo, T.%
\end{APACrefauthors}%
\unskip\
\newblock
\APACrefYearMonthDay{2018}{}{},
\newblock
\unskip
\newblock
\APACjournalVolNumPages{A\&A Rev.}{26}{1}{4}.
\newblock
\begin{APACrefDOI} \doi{10.1007/s00159-018-0109-x} \end{APACrefDOI}
\PrintBackRefs{\CurrentBib}

\bibitem [\protect \citeauthoryear {%
Moss%
\ \protect \BOthers {.}}{%
Moss%
\ \protect \BOthers {.}}{%
{\protect \APACyear {2017}}%
}]{%
Moss2017}
\APACinsertmetastar {%
Moss2017}%
\begin{APACrefauthors}%
Moss, V\BPBI A.%
, Allison, J\BPBI R.%
, Sadler, E\BPBI M.%
\ et al.\end{APACrefauthors}%
\unskip\
\newblock
\APACrefYearMonthDay{2017}{}{},
\newblock
\unskip
\newblock
\APACjournalVolNumPages{MNRAS}{471}{3}{2952--2973}.
\newblock
\begin{APACrefDOI} \doi{10.1093/MNRAS/STX1679} \end{APACrefDOI}
\PrintBackRefs{\CurrentBib}

\bibitem [\protect \citeauthoryear {%
{Ochsenbein}%
, {Bauer}%
\BCBL {}\ \BBA {} {Marcout}%
}{%
{Ochsenbein}%
\ \protect \BOthers {.}}{%
{\protect \APACyear {2000}}%
}]{%
Vizierpaper}
\APACinsertmetastar {%
Vizierpaper}%
\begin{APACrefauthors}%
{Ochsenbein}, F.%
, {Bauer}, P.%
\BCBL {}\ \BBA {} {Marcout}, J.%
\end{APACrefauthors}%
\unskip\
\newblock
\APACrefYearMonthDay{2000}{{\APACmonth{04}}}{},
\newblock
\unskip
\newblock
\APACjournalVolNumPages{\aaps}{143}{}{23-32}.
\newblock
\begin{APACrefDOI} \doi{10.1051/aas:2000169} \end{APACrefDOI}
\PrintBackRefs{\CurrentBib}

\bibitem [\protect \citeauthoryear {%
O'Dea%
}{%
O'Dea%
}{%
{\protect \APACyear {1998}}%
}]{%
ODea1998}
\APACinsertmetastar {%
ODea1998}%
\begin{APACrefauthors}%
O'Dea, C\BPBI P.%
\end{APACrefauthors}%
\unskip\
\newblock
\APACrefYearMonthDay{1998}{}{},
\newblock
\unskip
\newblock
\APACjournalVolNumPages{PASP}{110}{747}{493--532}.
\newblock
\begin{APACrefDOI} \doi{10.1086/316162} \end{APACrefDOI}
\PrintBackRefs{\CurrentBib}

\bibitem [\protect \citeauthoryear {%
O'Dea%
\ \BBA {} Saikia%
}{%
O'Dea%
\ \BBA {} Saikia%
}{%
{\protect \APACyear {2021}}%
}]{%
ODea2021}
\APACinsertmetastar {%
ODea2021}%
\begin{APACrefauthors}%
O'Dea, C\BPBI P.%
\BCBT {}\ \BBA {} Saikia, D\BPBI J.%
\end{APACrefauthors}%
\unskip\
\newblock
\APACrefYearMonthDay{2021}{}{},
\newblock
\unskip
\newblock
\APACjournalVolNumPages{A\&A Rev.}{29}{1}{1--109}.
\newblock
\begin{APACrefDOI} \doi{10.1007/s00159-021-00131-w} \end{APACrefDOI}
\PrintBackRefs{\CurrentBib}

\bibitem [\protect \citeauthoryear {%
Ostorero%
\ \protect \BOthers {.}}{%
Ostorero%
\ \protect \BOthers {.}}{%
{\protect \APACyear {2016}}%
}]{%
Ostorero2016}
\APACinsertmetastar {%
Ostorero2016}%
\begin{APACrefauthors}%
Ostorero, L.%
, Morganti, R.%
, Diaferio, A.%
, Siemiginowska, A.%
, Stawarz, L.%
, Moderski, R.%
\BCBL {}\ \BBA {} Labiano, A.%
\end{APACrefauthors}%
\unskip\
\newblock
\APACrefYearMonthDay{2016}{}{},
\newblock
\unskip
\newblock
\APACjournalVolNumPages{Astron. Nachrichten}{337}{1-2}{148--153}.
\newblock
\begin{APACrefDOI} \doi{10.1002/asna.201512284} \end{APACrefDOI}
\PrintBackRefs{\CurrentBib}

\bibitem [\protect \citeauthoryear {%
Predehl%
\ \protect \BOthers {.}}{%
Predehl%
\ \protect \BOthers {.}}{%
{\protect \APACyear {2021}}%
}]{%
Predehl2021}
\APACinsertmetastar {%
Predehl2021}%
\begin{APACrefauthors}%
Predehl, P.%
, Andritschke, R.%
, Arefiev, V.%
\ et al.\end{APACrefauthors}%
\unskip\
\newblock
\APACrefYearMonthDay{2021}{}{},
\newblock
\unskip
\newblock
\APACjournalVolNumPages{A\&A}{647}{}{12}.
\newblock
\begin{APACrefDOI} \doi{10.1051/0004-6361/202039313} \end{APACrefDOI}
\PrintBackRefs{\CurrentBib}

\bibitem [\protect \citeauthoryear {%
Siemiginowska%
\ \protect \BOthers {.}}{%
Siemiginowska%
\ \protect \BOthers {.}}{%
{\protect \APACyear {2016}}%
}]{%
Siemiginowska2016}
\APACinsertmetastar {%
Siemiginowska2016}%
\begin{APACrefauthors}%
Siemiginowska, A.%
, Sobolewska, M.%
, Migliori, G.%
, Guainazzi, M.%
, Hardcastle, M.%
, Ostorero, L.%
\BCBL {}\ \BBA {} Stawarz, {\L}.%
\end{APACrefauthors}%
\unskip\
\newblock
\APACrefYearMonthDay{2016}{}{},
\newblock
\unskip
\newblock
\APACjournalVolNumPages{ApJ}{823}{1}{57}.
\newblock
\begin{APACrefDOI} \doi{10.3847/0004-637x/823/1/57} \end{APACrefDOI}
\PrintBackRefs{\CurrentBib}

\bibitem [\protect \citeauthoryear {%
Snellen%
\ \protect \BOthers {.}}{%
Snellen%
\ \protect \BOthers {.}}{%
{\protect \APACyear {1998}}%
}]{%
Snellen1998}
\APACinsertmetastar {%
Snellen1998}%
\begin{APACrefauthors}%
Snellen, I\BPBI A\BPBI G.%
, Schilizzi, R\BPBI T.%
, {De Bruyn}, A\BPBI G.%
, Miley, G\BPBI K.%
, Rengelink, R\BPBI B.%
, R{\"{o}}ttgering, H\BPBI J.%
\BCBL {}\ \BBA {} Bremer, M\BPBI N.%
\end{APACrefauthors}%
\unskip\
\newblock
\APACrefYearMonthDay{1998}{}{},
\newblock
\unskip
\newblock
\APACjournalVolNumPages{Astron. Astrophys. Suppl. Ser.}{131}{}{435--449}.
\PrintBackRefs{\CurrentBib}

\bibitem [\protect \citeauthoryear {%
Snellen%
\ \protect \BOthers {.}}{%
Snellen%
\ \protect \BOthers {.}}{%
{\protect \APACyear {2000}}%
}]{%
Snellen2000}
\APACinsertmetastar {%
Snellen2000}%
\begin{APACrefauthors}%
Snellen, I\BPBI A\BPBI G.%
, Schilizzi, R\BPBI T.%
, Miley, G\BPBI K.%
, de Bruyn, A\BPBI G.%
, Bremer, M\BPBI N.%
\BCBL {}\ \BBA {} Rottgering, H\BPBI J\BPBI A.%
\end{APACrefauthors}%
\unskip\
\newblock
\APACrefYearMonthDay{2000}{}{},
\newblock
\unskip
\newblock
\APACjournalVolNumPages{MNRAS}{319}{2}{445--456}.
\newblock
\begin{APACrefDOI} \doi{10.1046/j.1365-8711.2000.03935.x} \end{APACrefDOI}
\PrintBackRefs{\CurrentBib}

\bibitem [\protect \citeauthoryear {%
{Tingay}%
\ \protect \BOthers {.}}{%
{Tingay}%
\ \protect \BOthers {.}}{%
{\protect \APACyear {2015}}%
}]{%
Tingay2015}
\APACinsertmetastar {%
Tingay2015}%
\begin{APACrefauthors}%
{Tingay}, S\BPBI J.%
, {Macquart}, J\BPBI P.%
, {Collier}, J\BPBI D.%
\ et al.\end{APACrefauthors}%
\unskip\
\newblock
\APACrefYearMonthDay{2015}{}{},
\newblock
\unskip
\newblock
\APACjournalVolNumPages{\aj}{149}{2}{74}.
\newblock
\begin{APACrefDOI} \doi{10.1088/0004-6256/149/2/74} \end{APACrefDOI}
\PrintBackRefs{\CurrentBib}

\bibitem [\protect \citeauthoryear {%
Vink%
, Snellen%
, Mack%
\BCBL {}\ \BBA {} Schilizzi%
}{%
Vink%
\ \protect \BOthers {.}}{%
{\protect \APACyear {2006}}%
}]{%
Vink2006}
\APACinsertmetastar {%
Vink2006}%
\begin{APACrefauthors}%
Vink, J.%
, Snellen, I.%
, Mack, K\BHBI H.%
\BCBL {}\ \BBA {} Schilizzi, R.%
\end{APACrefauthors}%
\unskip\
\newblock
\APACrefYearMonthDay{2006}{}{},
\newblock
\unskip
\newblock
\APACjournalVolNumPages{MNRAS}{367}{3}{928--936}.
\newblock
\begin{APACrefDOI} \doi{10.1111/j.1365-2966.2006.10036.x} \end{APACrefDOI}
\PrintBackRefs{\CurrentBib}

\bibitem [\protect \citeauthoryear {%
Worrall%
\ \protect \BOthers {.}}{%
Worrall%
\ \protect \BOthers {.}}{%
{\protect \APACyear {2004}}%
}]{%
Worrall2004}
\APACinsertmetastar {%
Worrall2004}%
\begin{APACrefauthors}%
Worrall, D\BPBI M.%
, Hardcastle, M\BPBI J.%
, Pearson, T\BPBI J.%
\ et al.\end{APACrefauthors}%
\unskip\
\newblock
\APACrefYearMonthDay{2004}{}{},
\newblock
\unskip
\newblock
\APACjournalVolNumPages{MNRAS}{347}{2}{632--644}.
\newblock
\begin{APACrefDOI} \doi{10.1111/J.1365-2966.2004.07243.X} \end{APACrefDOI}
\PrintBackRefs{\CurrentBib}

\bibitem [\protect \citeauthoryear {%
Yoon%
\ \protect \BOthers {.}}{%
Yoon%
\ \protect \BOthers {.}}{%
{\protect \APACyear {2020}}%
}]{%
Yoon2020}
\APACinsertmetastar {%
Yoon2020}%
\begin{APACrefauthors}%
Yoon, H.%
, Sadler, E.%
, Allison, J.%
\ et al.\end{APACrefauthors}%
\unskip\
\newblock
\APACrefYearMonthDay{2020}{}{},
\newblock
\unskip
\newblock
\APACjournalVolNumPages{AAS}{236}{}{322.05}.
\PrintBackRefs{\CurrentBib}

\end{thebibliography}

\section*{Author Biography}
\begin{biography}{\includegraphics[width=60pt,height=70pt]{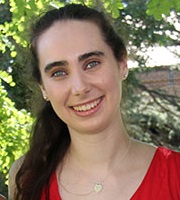}}{\textbf{Emily Kerrison} is an Honours student at the University of Sydney and CSIRO Space \& Astronomy. Her thesis is on the connection between \mbox{H\,{\sc I}} and X-ray absorption in obscured AGN, using data from the Australian SKA Pathfinder (ASKAP) and eROSITA.}.
\end{biography}

\end{document}